\documentstyle[aps,pra,epsf,twocolumn]{revtex}
\begin{document}
\draft
\twocolumn[\hsize\textwidth\columnwidth\hsize\csname @twocolumnfalse\endcsname
\title{On the zero temperature critical point in heavy fermions}
\author{Mucio A.Continentino}
\address{Instituto de Fisica,
Universidade Federal Fluminense \protect\\
Campus da Praia Vermelha, Niter\'oi, 24.210-340, RJ, Brasil \protect\\
and \protect\\
National High Magnetic Field Laboratory,
Florida State University \protect\\
1800 E.Paul Dirac Dr., Tallahassee, Florida, 32306, USA}
\date{\today}
\maketitle
\begin{abstract}
We generalize the  scaling theory of heavy fermions
for the case the shift exponent describing the critical N\'eel
line is different from the crossover exponent characterizing
the coherence line. We obtain the properties
of the non-Fermi liquid system at the critical point and in particular
the electrical resistivity.
We study violation of hyperscaling in the Fermi liquid regime
below the coherence line where the properties of heavy fermion
systems are described by mean-field exponents.
\end{abstract}

]

\section{Introduction}
Many of the physical properties of heavy fermions are determined
by the proximity of these systems to a zero temperature critical point
\cite{mucio1}.
The existence of a zero temperature phase transition, 
from a phase with long range magnetic order to a non-magnetic state,
allows for a formulation of a scaling theory which gives a very satisfactory
description of these materials \cite{mucio1}. This scaling theory is most
easily
formulated within the context of the Kondo lattice model which emphasizes the
importance of spin fluctuations neglecting charge fluctuations. This model has
two relevant energy scales, namely the bandwidth of the large conduction
band $W$, and the interaction $J$ between the local $f-electrons$ and the
conduction electrons in the wide conduction band. At $T=0$ the relevant
quantity is the ratio $J/W$ and the zero temperature phase transition occurs at
$(J/W)_c$
which is different from zero for $d \ge d_L$, the lower critical
dimension (probably $d_L =2$). For finite temperatures
the phase diagram for $d > d_L$, where a long 
range ordered magnetic phase exits at finite temperatures,
is shown in Figure 1.

This magnetic phase, generally of the antiferromagnetic type
occurs for $(J/W) < (J/W)_c$ and below a critical N\'eel line.

For $(J/W) > (J/W)_c$ and finite temperatures we pointed out
the existence of a crossover line, which we identified as
the {\em coherence line} and marks the onset,  with decreasing
temperature, of a renormalized Fermi-liquid regime on this
non-critical part of the phase diagram \cite{mucio1}.
This physical interpretation of the crossover line 
allows to recognize immeaditely  that a system just at the
critical point, i.e., with $(J/W) = (J/W)_c$,  should behave
as a non-Fermi liquid since it does not cross the coherence
line and consequently does not enter the Fermi liquid (FL)
region of the phase diagram \cite{mucio1,lohneysen,mucio2}.
In this paper we obtain the thermodynamic properties of this
critical system. We also discuss and clarify the question of
breakdown of  hyperscaling below the coherence line.

\begin{figure}[h]
\epsfxsize=3.5in  \epsfbox{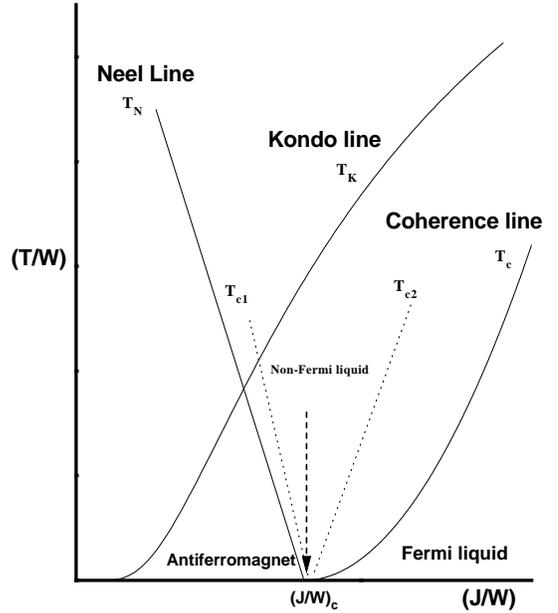}
\caption{The phase diagram of heavy fermions close to the zero temperature
critical point (schematic). In the present case the shift exponent  of the
N\'eel line $\psi =1$
is different from the crossover exponent $\nu z > 1$
characterizing the coherence line $T_c$.
The crossover lines $T_{c1}$ and $T_{c2}$ are both governed by the
same shift exponent of the N\'eel line \protect\cite{pfeuty}.
Below $T_c$ hyperscaling is violated and the thermodynamic properties are
governed by
mean-field exponents. }
\label{fig1}
\end{figure}

\section{Renormalization Group Equations}

The scaling results and the equations for the N\'eel and
coherence lines can be obtained considering the expansion of
renormalization group (RG) equations close to the zero temperature
fixed point at $(J/W)_c$. These expansions have a general character 
and can be written as,
\begin{equation}
K_{n+1} = K_c  +   b^{1/\nu}(K_n - K_c) -  (T/J)_n
\end{equation}
\begin{equation}
{\left(\frac{T}{J}\right)}_{n+1} = b^{z}{\left(\frac{T}{J}\right)}_n
\end{equation}
The first equation describes the renormalization of the ratio $K = (J/W)$
close to the  critical point $K_c =(J/W)_c$ under a change in length scale
by a factor $b$. The exponent $\nu$ is the correlation length exponent.
Eq. 2 describes the renormalization of the temperature close to
the zero temperature critical point. Since temperature is a parameter
its scaling is controlled by that of the coupling constant $J$ which at
the critical point scales  according to the
dynamical exponent $z$ \cite{mucio2}. In Eq. 1, temperature enters linearly
which
corresponds to the lowest order term in an analytic expansion
in powers of  $(T/J)$.
The solution of these recursion relations to a length scale $l =
b^n$ is given by:
\begin{equation}
K_l = K_c + l^{1/\nu}\left(K - K_c -  a(T/J)\right) + al^z\left(T/J\right)
\end{equation}
where $a$ is a constant.  Since $l$ is arbitrary we iterate until
$$ l^{1/\nu}\left( K - K_c -  a(T/J)\right) = 1$$
and this length scale defines the correlation length
$$l = \xi = {\vert K - K_c -  a(T/J) \vert}^{- \nu}= {\vert j -j_c(T) \vert}^{-
\nu}$$
where $j =(J/W)$ and  $j_c(T) = (J/W)_c +  a(T/J)$. 
At this length scale, substituting $l$ for $\xi$ in Eq.3, we obtain
$$K_{\xi} = K_c + 1 +  \frac{a(T/J)}{{\vert j - j_c(T) \vert}^{\nu z}}$$
The line $T_c/J = {\vert j - j_c(T) \vert}^{\nu z}$ in the non-critical
side of the phase diagram is a crossover line since $K_{\xi}$ and
consequently all functions of this ratio have different asymptotic
behaviors for $T >> T_c$ and $T << T_c$. 
It is governed by the exponent $\nu z$
and is identified as the coherence line in Figure 1. On the other hand,
at the line $j =j_c(T)$ or $T_N/J =  \frac{1}{a} \vert  j - j_c(T=0) \vert$
the correlation length diverges and this line can be identified with
the N\'eel line in the critical
region, i.e., for $(J/W) < (J/W)_c$.
If we define a {\em shift exponent \/} $\psi$ through the
relation $T_N/J \propto {\vert j -j_c \vert}^{\psi}$ \cite{pfeuty,millis}, the
present 
approach leads to $\psi = 1$ independent of the
crossover exponent $\nu z$. The critical line vanishes linearly
close to the zero temperature critical point which  is due
to the fact that in Eq.1 the lowest order analytic term in
temperature is a linear contribution.

\section{General Scaling Relations}

The general recursion relations, Eqs.1 and 2 give rise to
the following scaling form for the singular part of the free energy
density close to the zero temperature critical point \cite{mucio1}
\begin{equation}
f \propto {\vert \delta \vert}^{2- \alpha} f_0 [ T/T_c ]
\end{equation}
where $\delta = j - j_c(T=0)$ and $T_c \propto {\vert \delta \vert}^{\nu z}$.
In this section we consider the scaling properties of physical quantities in
the non-critical side of the phase diagram, i.e., for $(J/W) > (J/W)_c$. We
shall
neglect the temperature dependence of $j_c(T)$ which is
valid as long as $\nu z > \psi$ and $T \leq T_c$.
At $T=0$ , $f_0 (0) = constant$ and the above expression
gives the singular part of the ground state free energy density
close to the transition ($\delta=0$ defines the critical point).
In fact at $T=0$, Eq.4  defines the critical exponent $\alpha$.
The exponents $\alpha$, $\nu$ and $z$ defined above are associated
with the zero temperature unstable fixed point governing the quantum
phase transition \cite{mucio1}.

The existence of a FL regime for $T << T_c$ , as found 
experimentally, allows to take a Sommerfeld type of expansion for the
scaling function of the free energy density in this temperature
range \cite{mucio1}. We get
\begin{equation}
f \propto {\vert \delta \vert}^{2- \alpha} \{ a_1 + a_2 ( \frac{T}{T_c} )^2 +
a_3 ( \frac{T}{T_c} )^4 + \cdots  \}
\end{equation}
where $a_1$, $a_2$ and $a_3$ are constants.
The specific heat for $T << T_c$ is then given by
\begin{equation}
C \propto T \frac{ \partial^2 f}{ \partial T^2} \propto {\vert \delta 
\vert}^{2 - \alpha - 2 \nu z} T
\end{equation}
which shows the characteristic Fermi-liquid linear temperature
dependence with an effective thermal
mass $m_T \propto C/T  \propto  {\vert \delta  \vert}^{2 - \alpha -2 \nu z}$.

The scaling expression   for the free energy can be 
easily generalized to take into account the presence  of a  field
conjugate to the order parameter, in the present case a staggered field $H$ and
of a uniform
magnetic field $h$ \cite{mucio1}, we get
\begin{equation}
f \propto {\vert \delta \vert}^{2-\alpha}F \left[ \frac{T}{{\vert
\delta \vert}^{\nu z}}, \frac{H}{{\vert \delta \vert}^{\beta + \gamma}},
\frac{h}{{\vert \delta \vert}^{\phi_h}} \right]
\end{equation}
where $\nu$ and $z$ are the correlation length and dynamic exponents as
before, $\beta$ and $\gamma$ are standard critical exponents which
describe the critical behavior of the staggered magnetization and
susceptibility respectively.  $\phi_h$ is a new exponent associated
with the uniform magnetic field $h$. These exponents obey standard
scaling relations as $\alpha + 2 \beta + \gamma =2$ but the hyperscaling
relation is
modified  due to the quantum character of the critical point and is now given
by
\begin{equation}
2 - \alpha = \nu (d+z)
\end{equation}
where d is the dimensionality of the system
\cite{mucio1}.
{}From  Eq.7 we can deduce the scaling properties of the physical
quantities of interest like
the low temperature
($T << T_c$) field-dependent uniform magnetization
$m(h) \propto (\partial f/ \partial h)
\propto {\vert \delta \vert}^{2 - \alpha - \phi_h} f_m(h/h_c)$ with
$h_c \propto {\vert \delta \vert}^{\phi_h}$,
the uniform, low field,
susceptibility, $\chi_0 \propto {\partial}^2 f/ \partial h^2 \propto {\vert
\delta \vert}^{2 - \alpha - 2 {\phi}_h}f_S(T/T_c)$,
etc..

The scaling theory formulated above is completely general and any
description of heavy fermions in terms of a zero temperature phase
transition should fall within this approach. The crucial question is
of course to determine the critical exponents, i.e., the universality
class of the zero temperature critical point.  It is interesting
to point out that for $d =1 < d_L$,  $(J/W)_c = 0$, but there is  still  a
crossover line for $(J/W) > 0$ which is given by $k_B T_c \propto W
e^{\frac{-1}{J/W}}$,
an expression similar to  that of the Kondo line \cite{mucio1}.

\section{Gaussian Theories}

Eq. 6 for the specific heat  in the Fermi liquid regime ($T << T_c$)
can be written, using the modified  hyperscaling
relation, Eq.8,  as
\begin{equation} 
m_T \propto \gamma = C/T \propto {\vert \delta \vert}^{\nu (d-z)}
\end{equation}
For dimension $d=3$ and in the case the dynamic exponent $z=3$,
which is generally associated with  ferromagnetic fluctuations 
\cite{hertz} this gives essentially the result $\gamma 
\propto Ln \vert \delta \vert $ 
although the scaling theory
cannot reproduce the  logarithmic
singularity. This logarithmic dependence has the 
same origin of
the paramagnon mass
enhancement of a nearly ferromagnetic system obtained
originally by Doniach and Engelsberg \cite{doniach} for the Hubbard model. In
this case
$\vert \delta \vert = \vert (U/W) - (U/W)_c \vert$ where $U$ is 
the Coulomb repulsion, $W$ the bandwidth and $(U/W)_c$ 
is the critical ratio  \cite{doniach} at which the $T=0$ ferromagnetic
transition occurs.

The case of  antiferromagnetic fluctuations characterized by $z=2$
can be considered going to higher order in the Sommerfeld expansion of the free
energy.
Again using  hyperscaling, i.e.,  Eq.8,
we get
\begin{equation}
\gamma = C/T  \propto a_2 \vert \delta \vert^{\nu(d-z)} + a_3
\vert \delta \vert^{ \nu (d-3z)} T^2
\end{equation}
In three dimensions with $z=2$ \cite{hertz} \cite{millis} this yields
\begin{equation}
C/T \propto a_2 { \vert \delta \vert }^{1/2} + a_3 \frac{T^2}{ { \vert \delta 
\vert }^{3/2} }
\end{equation}
for $T << T_c$ where we took the Gaussian exponent $\nu = 1/2$. This result
coincides with that of the Gaussian theory of Millis \cite{millis} for
a nearly antiferromagnetic Fermi liquid and relies on the validity of
the hyperscaling relation, Eq.8, below $T_c$.
In fact the exponent $\alpha$ in Gaussian theories of zero temperature magnetic
phase
transitions is determined by the modified hyperscaling relation i.e. $\alpha =
\frac{4 - (d+z)}{2}$,
where we used the Gaussian value
for the correlation length exponent  $\nu =1/2$.
Consequently in these theories hyperscaling is automatically  satisfied.
Notice that for  $d+z > 4$ the Gaussian exponent $\alpha < 0$ implying
that singularities in these theories are weaker
than those obtained in a mean-field approach where the exponent $\alpha$ is
fixed at
$\alpha = 0$
for $d+z > d_c$. Here  $d_c = 4$ is the upper critical dimension 
for these magnetic transitions \cite{hertz,toulouse}. However the mean field
exponents
$\alpha =0$ and $\nu = 1/2$ violate hyperscaling above $d_c$, i.e.,
do not satisfy Eq.8 for $d+z > 4$.

Let us consider again the
enhancement of the 
thermal mass  for an itinerant three dimensional nearly
ferromagnetic system on the light of the discussion above.
In the general scaling approach we have, $ m_T 
\propto {\vert \delta \vert}^{2- \alpha - 2\nu z}$,
which is obtained without using the hyperscaling relation.
Using the mean-field exponents $\alpha = 0$, $\nu = 1/2$ and $z = 3$,
which violate Eq.8 for $d=3$,
we get 
$m_T \propto {\vert (U/W) - (U/W)_c \vert}^{-1} = { \vert \delta \vert}^{-1}$.

Since the mean-field susceptibility exponent $\gamma = 1$,  we find
$\chi_0 \propto {\vert \delta \vert}^{-1}$ where $\chi_0$ is the
limiting, $T \rightarrow 0$, uniform susceptibility. Consequently
the ratio 
$\chi_0 / m_T = constant$
independent of $(U/W)$, within mean-field,  differently from
the paramagnon or Gaussian results  which yield
$m_T \propto Ln \vert \delta \vert $, as we saw previously
and consequently to a diverging ratio $\chi_0 / m_T$ as $\vert
\delta \vert \rightarrow 0$ \cite{doniach} \cite{millis}.

It is interesting to compare these different predictions with measurements
at low temperatures of the electronic and magnetic Gruneisen parameters
of bulk {\em Palladium}, the prototype of an itinerant $3-d$ nearly
ferromagnetic
system. These pressure  experiments in which the ratio $(U/W)$ is varied
due to it's dependence on volume $V$  yield \cite{mydosh}:
\[ \frac{ \partial Ln \left[ \chi_0/m_T \right]}{\partial LnV} = \]
\[ \frac{\partial Ln \chi_0}{\partial LnV} - 
\frac{\partial Ln m_T}{\partial LnV} = +2.8 - 2.2 = +0.6 \pm 0.5 \]
This {\em near cancellation \/} \cite{mydosh} implies
$\chi_0 / m_T \approx constant$ 
in good agreement with the mean-field scaling prediction. 
The analysis above points
out that the true critical behavior of the renormalized,
nearly magnetic, Fermi liquid below the coherence line is
governed by mean-field exponents.
Since the existence of paramagnons in nearly ferromagnetic systems is
well established \cite{valls} we expect that these Gaussian type of
fluctuations
become important above this line as we shall further
discuss.

The breakdown of hyperscaling in the Fermi liquid regime
{\em below the coherence line \/} can be
more specifically  discussed within  Hertz approach to quantum
magnetic phase transitions \cite{hertz}. In this  case it
can be seen how it arises. It
is due to the {\em dangerous irrelevant \/} quartic interaction
$u_0$ in the field operators,  for $d+z > 4$
\cite{hertz}.
The scaling form of the singular part of the $T=0$ free energy can be
written in this approach as
$f \propto {\vert \delta \vert}^{\nu(d+z)}F[u_0 {\vert \delta
\vert}^{\frac{d+z-4}{2}}]$
\cite{hertz,millis}.
The dangerous  irrelevant nature of $u_0$ is a consequence that
$F[x \rightarrow 0] \propto 1/x$.
This asymptotic behavior of the scaling function can be easily
obtained if we consider the expansion of
the free energy in the ordered magnetic
phase in terms of the order  parameter $\eta$, i.e.,
$f \propto {\eta}^2 + u_0 {\eta}^4 + \ldots$ and recall that
$\eta \propto (u_0)^{-1/2}$ in Landau theory \cite{toulouse}.
Using this result for $F[x \rightarrow 0]$
we get $f \propto {\vert \delta \vert}^{\nu (d+z)}/\left( u_0
{\vert \delta \vert}^{\frac{d+z-4}{2}} \right)$ which leads to
$f \propto {\vert \delta \vert}^2$ and consequently to the
mean-field exponent $\alpha = 0$
which violates Eq.8 for $d + z > 4$ since $\nu = 1/2$.

The argument presented above, for $T = 0$,  should hold below the
coherence line, i.e., for
$T < T_c$. For $T > T_c$, the interaction $u_0$ couples
to temperature giving rise to a new effective field
($u_0T$)  which is now {\em relevant \/} \cite{millis}.
Consequently  the previously discussed mechanism for violation of hyperscaling
associated with the dangerous irrelevance of $u_0$
is inoperative  for $T > T_c$.
Then, for $T > T_c$ and in
particular at $\delta = 0$,  there is no reason
to expect violation of hyperscaling and Gaussian
fluctuations should become dominant.
This is indeed indicated by the experiments
in heavy fermion systems  as we will discuss.

\section{The Heavy Fermion Fixed Point}

The analysis of experimental data on heavy fermions {\em below the coherence
line},
$T \leq T_c$,
has led to
the following empirical relations between the 
exponents associated with the 
zero temperature fixed point:  $2 - \alpha = \nu z$ and $\phi_h =
\nu z$ \cite{sachdev}, where $\phi_h$ controls the scaling of the 
uniform magnetic field \cite{mucio1,mucio2}. The former relation clearly
violates hyperscaling, Eq.8, for $d=3$
suggesting  that the fixed point governing the
physics of heavy fermions when approached from below the coherence
temperature ($T < T_c$) is characterized by
classical {\em mean-field exponents \/}
at least
in $d=3$. As we pointed out above
mean-field exponents,
contrary to  the Gaussian   ones, violate hyperscaling
above the upper critical dimension $d_c$, i.e., in the case of quantum
transitions,
for $d+z > d_c$ \cite{toulouse}.
Notice that the relations $2 - \alpha = \nu z$ and $\phi_h =
\nu z$, which lead to $2 - \alpha = \phi_h$ imply that the $T=0$ uniform
susceptibility  behaves
as $\chi_0 \propto {\vert \delta \vert}^{- \nu z}$
and the thermal mass,
$m_T \propto {\vert \delta \vert}^{- \nu z}$
as the critical point is approached
at zero temperature. This  behavior of $\chi_0$ shows
the
importance of ferromagnetic fluctuations and the reason for the enhanced
uniform susceptibility in heavy fermions.
Since the staggered susceptibility also diverges at the transition,
the zero temperature critical point which governs the physics of heavy
fermions
is in fact a multicritical point. This led us to consider that the heavy
fermion
fixed point is in the universality class of the classical tricritical
point with $\alpha = 1/2$,  $\nu =1/2$, $\gamma =1 $, ${\phi}_T = \nu
z/{\phi}_h = 1$
and furthermore with the dynamic exponent $z$
assuming the value $z=3$. This yields for the coherence line
the equation $T_c \propto {\vert \delta \vert}^{3/2}$.
Since $\nu z > 1$, this line rises from the $T=0$ axis
with zero derivative reducing the Fermi liquid region close
to the critical point for $(J/W) > (J/W)_c$. This may be
responsible for the ubiquity of non-Fermi liquid behavior  in
heavy fermions driven to a non-magnetic state by chemical or external pressure
\cite{budko,loidl}.
There are of course other possible scenarios as the nearly
antiferromagnetic Fermi liquid \cite{hertz,millis} with $z=2$ such that
$\nu z =1$.
For disordered systems one finds $z=4$
associated with diffusive modes \cite{hertz} and $\nu = 1/2$ such that $\nu z
=2$.

\section{Non-Fermi Liquid Behavior}
 
Fig.1 shows  the phase diagram of heavy fermions for
$\nu z > \psi = 1$ such that the critical line
vanishes linearly close to the
$T=0$ critical point
as observed  \cite{lohneysen}.
It is clear from this figure and our interpretation of the crossover line that
a system just at the critical point, i.e., with $(J/W) = (J/W)_c$,
$\vert \delta \vert =0$, does not
cross the coherence line and consequently does not enter the Fermi liquid
regime.
In  previous papers \cite{mucio1,mucio2} we have obtained the  properties of
such
a system under the assumption that $\psi = \nu z$ which is the
content of the so-called {\em generalized scaling hypothesis \/} \cite{fisher}.
The  experiments however indicate that this may not hold in heavy fermions
and in fact $\nu z > \psi =1$ \cite{lohneysen}.
We shall then calculate the  thermodynamic
properties of the system at the critical point for the
case $\nu z \ge \psi$. For this purpose we introduce a
set of critical exponents, which we distinguish by a {\em tilde \/}
and describe the singularities along the critical N\'eel
line \cite{til}.
It is clear that these {\em tilde \/} exponents are
different from those of the zero temperature critical
point. In the renormalization group language this is a consequence
that temperature is a {\em relevant field \/}  and consequently the
finite temperature transitions are governed by another, $T \ne 0$, fixed point.

\subsection{Free energy}

Let us consider the expression for the free energy density,
\begin{equation}
f \propto {\vert j - j_c(T) \vert}^{2  - \alpha}F[t]
\end{equation}
where 
\begin{equation}
t = \frac{T/J}{{\vert j - j_c(T) \vert}^{\nu z}}
\end{equation}
and 
\begin{equation}
j_c(T) - j_c(0) \propto T^{1/\psi}
\end{equation}
The scaling function $F[t]$ has the following asymptotic behaviors.
When $t \rightarrow 0$, $F[0] = constant$.
On the other hand for $j \rightarrow j_c(T)$, i.e.,
for $t \rightarrow \infty$ we should recover the critical
behavior of the  finite temperature antiferromagnetic transition,
then $F[t \rightarrow  \infty ] \propto t^x$
with $x = \frac{\tilde{\alpha} - \alpha}{\nu z}$ \cite{pfeuty}.
This guarantees that in this limit
$f \propto {\vert j - j_c(T) \vert}^{2 - \tilde{\alpha}}A(T)$ or
$f \propto {\vert T - T_N(j) \vert}^{2 - \tilde{\alpha}}A(T)$
where the amplitude $A(T) = T^{\frac{\tilde{\alpha}- \alpha}{\nu z}}$. 
At the critical point, i.e., $(J/W) = (J/W)_c$, we obtain
\begin{equation}
f \propto T^{\frac{2 - \tilde{\alpha}}{\psi} + \frac{\tilde{\alpha} -
\alpha}{\nu z}}
\end{equation}
Let us consider some particular cases.

{\em i}) $\psi = \nu z$.
In this case $f \propto T^{\frac{2 - \alpha}{\nu z}}$.
Using hyperscaling,
i.e., Eq.8, we find
a correction to the singular part of the free energy density
$f \propto T^{\frac{d+z}{z}}$ which yields for the specific heat
$C/T \propto T^{\frac{d-z}{z}}$. In particular for $d=z$ this yields
$C/T \propto LnT$.
On the other hand notice that using the empirical
mean-field relation $2 - \alpha = \nu z$ we get $f \propto T$
and the specific heat is
determined by analytic contributions to the free energy such that
$C \propto T$.

{\em ii}) $\psi \le \nu z$.
In this case the specific heat is generally given by,
$$ C/T \propto \frac{{\partial}^2 f}{\partial T^2} \propto
T^{\frac{(2 - \tilde{\alpha})( \nu z - \psi) + \psi ( 2 - \alpha - 2 \nu
z)}{\nu z \psi}}$$ 
Hyperscaling   yields
$$ C/T \propto \frac{{\partial}^2 f}{\partial T^2} \propto
T^{\frac{(2 - \tilde{\alpha})( \nu z - \psi) + \psi \nu( d -
z)}{\nu z \psi}}$$ 
for  $\psi=1$, $d = z =3$ and $\nu z = 3/2$ we get
$C/T \propto T^{\frac{2 - \tilde{\alpha}}{3}}$.
For $z=2$, such that $\nu z = \psi = 1$,  we obtain
$C/T \propto T^{1/2}$ as found previously by Millis
using a different approach \cite{millis}. In every case we should
consider an additional temperature independent regular
contribution to $C/T$.

The mean-field result is obtained
using the empirical relation $2 - \alpha = \nu z$,  we
get
$$C \propto T^{\frac{(2 - \tilde{\alpha})(\nu z - \psi)}{\nu z \psi}}$$
In the case $\psi = 1$, $\nu z =3/2$ and $\tilde{\alpha} \approx 0$ we get a
stronger than logarithmic divergence for the thermal mass, namely, $C/T \propto
T^{-1/3}$.
For $\psi = 1$, $\nu z =2$, which is the disordered case, we
get $ C/T \propto T^{-\tilde{\alpha}/2}$.
For $\tilde{\alpha}$ small  this gives essentially
$C/T \propto LnT$.

\subsection{Order Parameter Susceptibility}

We start with the scaling expression
$$ \chi_s \propto {\vert j - j_c(T) \vert}^{- \gamma} F(t)$$
which close to the critical N\'eel line becomes
$$ \chi_s \propto {\vert j - j_c(T) \vert}^{- \tilde{\gamma}}
T^{\frac{\tilde{\gamma} - \gamma}{\nu z}}$$
For $(J/W) = (J/W)_c$ we get
\begin{equation}
\chi_s \propto T^{- \tilde{\gamma} \left( \frac{1}{\psi} - \frac{1}{\nu z}
\right) - \frac{\gamma}{\nu z}}
\end{equation}
Taking $\psi =1$ and $\gamma =1$, appropriate for $d + z > d_c$, yields
$ \chi_s \propto T^{- \frac{2 + \tilde{\gamma}}{3}}$ for $\nu z =3/2$ and
$\chi_s \propto T^{-1}$ for $\nu z =1$.

\subsection{Uniform Susceptibility}

If we take into account that the uniform susceptibility does not diverge at
the critical N\'eel line  we  get the following expression for $\chi_0$ at
$(J/W)_c$,
\begin{equation}
\chi_0 \propto T^{\frac{2 - \alpha - 2 \phi_h}{\nu z}}
\end{equation}
Now using hyperscaling,
which is consistent with Gaussian exponents
and the relation ${\phi}_h = \nu z$ \cite{sachdev},  we
obtain
\begin{equation}
{\chi}_0 \propto T^{\frac{d - z}{z}}
\end{equation}
which, assuming $z=3$,  leads to
a constant or at most a weakly logarithmic  divergent susceptibility at
$(J/W)_c$. In this case  also
$dm/dh \propto Lnh$ or constant.

On the other hand taking  the dynamic exponent
$z=2$, appropriate for antiferromagnetic fluctuations,
yields,  $\chi_0 \propto T^{1/2}$, at $\delta = 0$,
eventually with an additional,
temperature independent, regular contribution. We also get $dm/dh \propto
h^{1/2}$ in this case. The above temperature dependence of $\chi_0$
has in fact been
observed \cite{lohneysen}.

Notice that using the mean-field empirical relation $2 - \alpha = \nu z$
and $\nu z  = \phi_h$ we find that at the
critical point $\chi_0 \propto T^{-1}$ which is the expected
mean-field result
for a N\'eel transition ocurring at $T=0$. This singular behavior
is quite different from that observed.
For the differential uniform susceptibility mean-field yields at the critical
point,
$\chi_h = dm/dh \propto {h}^{-1}$.

The behavior of the uniform
susceptibility at $\vert \delta \vert = 0$ is
relevant  as concerns the possibility of superconductivity
at this point.
In fact a  $\chi_0$ diverging with temperature  at the critical point
would appear  detrimental
for
superconductivity since
it signals the presence of unscreened net
magnetic moments.

\subsection{Nuclear Relaxation Time}

The nuclear relaxation time is given by \cite{bourbonnais}:
\begin{equation}
\frac{1}{T_1} = k_B T  \sum_q { \vert A_q \vert }^2 \frac{Im \chi (q,
{\omega}_N)}{{\omega}_N}
\end{equation}
where $A_q$ is the form factor and Im$\chi$ is the imaginary part of the
wavevector and frequency dependent susceptibility $\chi (q, \omega)$
of the system at the nuclear resonance  frequency ${\omega}_N$.
This expression has contribution from all wavevectors $q$.
Let us consider the contribution of wavevectors
near $q  =  Q_0$ the wavector characterizing the ordered
antiferromagnetic phase.
Using the dynamic scaling hypothesis \cite{halperin}
\begin{equation}
Im \chi (q, {\omega}_N)= {\chi}_{s}(T)D(q \xi  , {\omega}{\xi}^z)
\end{equation}
where $\xi$ is the correlation length and $z$ the dynamic exponent and
substituting in the previous expression we get \cite{bourbonnais}
\begin{equation}
\frac{1}{T_1}= k_B T \int{d^d q \frac{ {\chi}_{s}(T) D(q \xi , \omega
{\xi}^z) }{{\omega}_N}}
\end{equation}
which yields
\begin{equation}
\frac{1}{T_1} \propto T {\chi}_{s}(T) {\xi}^{z-d}
\end{equation}
Note that the quantity
${\xi}^{z-d}$
scales as the thermal mass, i.e.,  ${\xi}^{z-d} \propto m_T \propto C/T$
such that $\frac{1}{T_1} \propto {\chi}_{s}(T) C(T)$.
Consequently if
$C/T \propto LnT$
\cite{lohneysen}, we expect to find using the previous
results for $\chi_s$ at least a logarithmic diverging linewidth, i.e.,
$\frac{1}{T_1} \propto LnT$ close to the critical point
($\tilde{\gamma} \ge  1$ \cite{til}).

For completeness we  give the temperature dependence of the correlation
length at the critical point, we get, ${\xi}  \propto T^{- \tilde{\nu}/
\psi
+ ( \tilde{\nu} - \nu )/ \nu z}$.

We now return to the phase diagram of Fig.1.
The form of the scaling function Eq.12 together with Eqs. 13 and 14  allow  to
identify two additional crossover lines in this
phase diagram,
$T_{c1}$ and $T_{c2}$,
whenever $\psi \ne \nu z$ \cite{pfeuty}.
Both these lines are governed by the same exponent $\psi$ of the
critical N\'eel line \cite{pfeuty,millis}, i.e., $T_{c1} \propto {\vert j
\vert}^{\psi}$
and $T_{c2} \propto {\vert j \vert}^{\psi}$. The line to the left of $(J/W)_c$,
$T_{c1}$,
marks the onset of the classical regime where the singularities close
to the N\'eel line  are described
by the
{\em tilde exponents}. In the non-critical side of the phase diagram
the temperature dependence
of the physical quantities in the regime between $T_{c2}$ and the coherence
line
($T_c < T < T_{c2}$),
are obtained from the expressions deduced before for
$\delta = 0$ and taking $\psi = \nu z$
since $\psi$ is the relevant crossover exponent in this temperature
interval \cite{pfeuty}. In this region we expect hyperscaling to
hold and that Gaussian fluctuations
become
dominant.
The thermal mass for example
in this  temperature region
($T_c < T < T_{c2}$)
is given by
$C/T \propto LnT$, for $z=3$.

\subsection{Resistivity at the critical point}

The resistivity at the critical point can be easily obtained
generalizing previous calculations that consider the scattering
of conduction electrons by different type of bosons
\cite{lederer,mathon,rivier}.
The resistivity can be expressed, using the
fluctuation-dissipation theorem, in terms
of the dynamic susceptibility  $\chi (k, \omega)$ associated with
the elementary excitations \cite{lederer}.
In three dimensions it is given by \cite{rivier}
$$\rho_c = A \beta \int_{0}^{\infty}d \omega \frac{\omega}{\left[ e^{\beta
\omega} -1 \right]
\left[ 1 - e^{- \beta \omega} \right]} \times$$
\begin{equation}
\int_0^{2k_F} dk k^3 {\vert F_k \vert}^2
Im \chi (k, \omega)
\end{equation}
where $A = \frac{3}{{4 e \hbar K_F^3}^2}$, $\beta = \frac{1}{k_B T}$ and $F_k$
is a form
factor.
A similar expresion but with the integral in $k$ extending in the interval
$[0 , + \infty ]$ and weighted by a factor $k^2$ rather than $k^3$ is obtained
if the scattering of the electrons by the bosons does  not conserve momentum
\cite{rivier}.
This may occur in a disordered alloy in which case the boson propagator
should be averaged over all wavectors \cite{mucio3}. In this case there is
always a
finite residual resistivity.
In order to make contact with previous approaches \cite{lederer,mathon} we
consider, without loss of generality,  an explicit expression for the
imaginary part of the dynamic  susceptibility at the critical point
\begin{equation}
Im \chi (k, \omega) = {\chi}_{static} \frac{\tilde{\omega}}{{\tilde{\omega}}^2
+ {\Delta}^2}
\end{equation}
where $\tilde{\omega} = ( \omega / k^l )$, $\Delta = Dk^m$. Taking $F_k =
Ck^p$ and
making a double change of variables in the integrals for the
resistivity we obtain \cite{rivier}
\begin{equation}
\rho \propto T^{\frac{2p + 4 + l}{z}}
\end{equation}
and in the case momentum is not conserved
$${\Delta \rho}_{Alloy} \propto T^{\frac{2p + 3 + l}{z}}$$
where $z = l + m$ is the dynamic critical exponent.
We neglect the k-dependence of the form factor ($p=0$) and the
temperature dependence of ${\chi}_{static}$.
For $p = 0$, $l =1$ and $z=3$ this gives the
result $\rho  \propto T^{\frac{5}{3}}$ obtained by Mathon \cite{mathon} due
to critical paramagnons. In the same case but without momentum conservation
one finds at the critical point ${\Delta \rho}_{Alloy} \propto
T^{\frac{4}{3}}$.
For diffusive modes with $l =2$ and $z = 4$ we get, $\rho \propto T^{3/2}$
and without momentum conservation ${\Delta \rho}_{Alloy} \propto T^{5/4}$ at
the
critical point.

The case of antiferro-paramagnons, $l =0$, $z=2$ yields,
$\rho \propto T^2$ and ${\Delta \rho}_{Alloy} \propto T^{\frac{3}{2}}$.
In the former case, taking into account the temperature dependence of
${\chi}_{static} = {\chi}_s \propto T^{-1}$ obtained before,  we get a
{\em linear temperature dependent} resistivity at the
critical point as observed  \cite{lohneysen}.

It is interesting to consider the high temperature
behavior ($T >> T_c$)  of Eq. 23 for the resistivity away from the critical
point. Using the Kramers-Kronig relation we find
$$\rho \approx  Ak_B T \int_0^{2k_F} dk k^3 {\vert F_k \vert}^2
\chi_{static} (k)$$
This high temperature linear behavior is characteristic of alloys
where a crossover to a $T^2$ behavior at low temperatures is
also observed \cite{kaiser}.

\section{Discussion and Conclusions}

Using an expansion of  the renormalization group equations
close to the zero temperature fixed point  describing the physics
of heavy fermions we have obtained a critical  line which
vanishes linearly close to the zero temperature critical
point independently of the crossover exponent $\nu z$ \cite{primeira}.
We have then generalized  our previous scaling approach
for the case the shift exponent describing the   critical N\'eel line
is different from the crossover exponent of the coherence line.
In particular we have obtained the properties of the non-Fermi
liquid system at the critical point, i.e., for $(J/W) = (J/W)_c$.
In the general case that $\psi \ne \nu z$  we have to consider two new
crossover lines in the phase diagram of Figure 1,
both governed by the shift
exponent $\psi$ \cite{pfeuty}.
In the region between $T_{c1}$ and $T_{c2}$, which includes the
non-Fermi liquid regime above  the critical
point, the {\em tilde \/} exponents which describe the singularities
along the N\'eel line
appear explicitly in the temperature dependence of the physical
quantities besides those associated with the $T =0$ fixed point.
We have considered different scenarios or universality classes
for the $T=0$ transition.

Our previous analysis of the pressure dependence of several physical quantities
in different heavy fermion systems   below the
coherence temperature
led to    the relations $2 - \alpha = \nu z$ and
${\phi}_h = \nu z$.
The former equality is inconsistent with hyperscaling
implying that
in this region of the phase diagram hyperscaling
is violated and the physical properties are described by {\em mean-field \/}
exponents.
Due to the generality of the concept of {\em coherence line \/}
this result   is
also valid  for nearly ferromagnetic systems  as
our analysis of the pressure dependence of the Wilson ratio of $Pd$ has shown.

While mean-field exponents
are appropriate to describe the Fermi liquid regime
for $T < T_c$,
above the coherence  line and in
particular at the critical point hyperscaling is eventually restored.
Experimentally this is shown, in the case of heavy fermions,
by the saturation of the uniform
susceptibility at very low temperatures and the weak divergence
of the thermal mass at the critical point.
This can be obtained using hyperscaling while according to
mean-field $\chi_0$
is expected to diverge strongly  ($\chi_0 \propto T^{-1}$)
and $C/T$ has a stronger than logarithmic divergence.
In the context of the phenomenological scaling approach
it is not possible to determine {\em a  priori \/} the
validity of hyperscaling in the different regions of
the phase diagram.
However using
Hertz approach to quantum magnetic phase transitions
we have discussed how a mechanism for the  breakdown of hyperscaling
at  low temperatures, i.e., below
the coherence line,
in fact becomes  ineffective for $T > T_c$ and hyperscaling
is restored above this line.

We  have pointed  out the relevance of
ferromagnetic fluctuations
in heavy fermions as indicated, for example, by the enhanced
uniform susceptibility.
These fluctuations, which  are associated with the dynamic exponent $z=3$,
may be due to the existence of
ferromagnetically correlated planes which in turn order
antiferromagnetically. This type of correlations is often
found in metamagnetic systems \cite{haen}.
Our analysis has shown that in the case $\nu z \ne  \psi$ we do not recover
in general, for $z=3$,  the result $C/T \propto LnT$ at $\delta = 0$
but away from the critical point, i.e.,
for $T_c < T < T_{c2}$.
This region extends down to very low temperatures close to $\delta = 0$
since $T_c \propto {\vert \delta \vert}^{3/2}$ making it
difficult to identify the critical point.
Also away from $\delta = 0$ and  $T > T_c$,
it is easier to
obtain a linear temperature dependent resistivity as outlined in section $6.5$.

We found that taking  the dynamic exponent
$z=2$ yields the observed temperature dependence of
the uniform susceptibility  at $\delta =0$  \cite{lohneysen}.
In this case, even without considering the existence of
{\em hot lines}  in the Fermi surface \cite{rice},  we could obtain
a linear temperature dependent resistivity at $\delta = 0$  \cite{aliev}.
It is clear from our previous remarks that
the case of $z=2$ is also the most favorable
for the appearance of superconductivity at the critical point
\cite{super}.

We hope further experiments are carried out to distinguish
between the different scenarios and determine unambiguously the
universality class of the zero temperature heavy fermion
fixed point. The general scaling theory formulated above provides
a powerful tool to acomplish this task.

The scaling theory has proved to be useful to describe
the properties of $Ce$ based heavy fermions. However in the case of  $U$
compounds, probably
due to the extension of the $5-f$ orbitals,
new phenomena appear as the coexistence of long range
magnetic order and superconductivity \cite{steglich} which would be interesting
to incorporate in the present approach.
On the other hand
there is a family of materials based
on $Uranium$ which exhibit many of the non-Fermi
liquid properties discussed above  \cite{andraka}  \cite{tsvelik}. Whether
these systems,
which have high residual resistivities, may be described by the
ideas discussed here, possibly with differents exponents or within a {\em
single
impurity \/} scenario \cite{vlad} remains to be investigated.

\acknowledgements
I would like to thank Profs. F. Steglich, H. v. Lohneysen, A. Loidl,
Raimundo R. dos Santos  and
Drs. E. Miranda, N. Bonesteel, and V. Dobrasavljevic
for useful discussions. I am particularly
thankful for Dr. E. Miranda for a critical
reading of the manuscript. I also thank CNPq-Brasil
for partial financial support.







\begin{references}
\bibitem{primeira} Notice that if the $T=0$ transition is first order then
$dT_N/dP \rightarrow - \infty$ as $T_N  \rightarrow 0$ or $P \rightarrow P_c$
where $P_c$ is the critical pressure; Samara G.A., Terauchi H., Phys.Rev.Lett.
{\bf 59}, 347 (1987).
\end{references}
\end{document}